\documentclass[prd,twocolumn,showpacs,amsmath,amssymb,widetext,floatfix]{revtex4-1}
\usepackage{graphicx}
\usepackage{dcolumn}
\usepackage{bm}
\usepackage{epsfig}
\usepackage{color}
\usepackage{hyperref}

\hypersetup{
    colorlinks=true,
    linkcolor=red,
    citecolor=blue,
}

\def\fun#1#2{\lower3.6pt\vbox{\baselineskip0pt\lineskip.9pt
  \ialign{$\mathsurround=0pt#1\hfil##\hfil$\crcr#2\crcr\sim\crcr}}}
\def\simgt{\mathrel{\lower0.6ex\hbox{$\buildrel {\textstyle >}
 \over {\scriptstyle \sim}$}}}
\def\simlt{\mathrel{\lower0.6ex\hbox{$\buildrel {\textstyle <}
 \over {\scriptstyle \sim}$}}}

\input epsf

\newcommand{\mnras}{MNRAS}
\newcommand{\apjl}{ApJL}
\newcommand{\aj}{AJL}
\newcommand{\aap}{A$\&$A}

\newcommand{\gth}{G_\Theta} 

\def\be{\begin{equation}}
\def\ee{\end{equation}}
\def\ba{\begin{eqnarray}}
\def\ea{\end{eqnarray}}

\def\nn{\nonumber}

\newcommand{\hompc}{\,h\,{\rm Mpc}^{-1}}
\newcommand{\mpcoh}{\,h^{-1}\,{\rm Mpc}}

\newcommand{\zeff}{z_{\rm eff}} 
\newcommand{\om}{\Omega_m} 
\newcommand{\jcap}{J. Cosmology Astropart. Phys.}

\begin{document}

\preprint{}

\title{Cosmological Constraints from the Anisotropic Clustering Analysis using BOSS DR9}
 
\author{Eric V.\ Linder$^{1,2,3}$, MinJi Oh$^{2,4}$, Teppei Okumura$^3$, 
Cristiano G.\ Sabiu$^5$, Yong-Seon Song$^{2,4}$} 
\email{authors by alphabetical order}
\email{ysong@kasi.re.kr}
\affiliation{$^1$Berkeley Lab and Berkeley Center for Cosmological Physics, 
University of California, Berkeley, CA 94720, USA} 
\affiliation{$^2$Korea Astronomy and Space Science Institute, Daejeon 305-348, 
Korea} 
\affiliation{$^3$Institute for the Early Universe WCU, Ewha Womans University, 
Seoul 120-750, Korea} 
\affiliation{$^4$University of Science and Technology, Daejeon 305-333, Korea}
\affiliation{$^5$Korea Institute for Advanced Study, Dongdaemun-gu, Seoul 
130-722, Korea}  

\date{\today}

\begin{abstract}
Our observations of the Universe are fundamentally anisotropic, with data 
from galaxies separated transverse to the line of sight coming from the same 
epoch while that from galaxies separated parallel to the line of sight coming 
from different times.  Moreover, galaxy velocities along the line of sight 
change their redshift, giving redshift space distortions.  We perform a full 
two-dimensional anisotropy analysis of galaxy clustering data, fitting in a 
substantially model independent manner the angular diameter distance $D_A$, 
Hubble parameter $H$, and growth rate $d\delta/d\ln a$ without assuming a 
dark energy model.  The results demonstrate consistency with $\Lambda$CDM 
expansion and growth, hence also testing general relativity. We also point out the interpretation dependence of the 
effective redshift $z_{\rm eff}$, and its cosmological impact for next 
generation surveys. 
\end{abstract}



\maketitle

\section{Introduction}

Large volume galaxy redshift surveys are mapping large scale structure in the 
universe, measuring the three dimensional positions of millions of galaxies. 
This data teaches us not only the statistics of clustering but can be used 
to measure cosmic distances and growth, and constrain cosmological models.  
The third dimension of the data, the redshift, allows investigation of two 
effects: the role of cosmic expansion in distinguishing transverse and radial 
distances, and the role of peculiar velocities, measured through their induced 
redshift space anisotropy, as a probe of the cosmic structure growth rate. 

Redshift space distortions (RSD) are induced by the large scale velocity flow 
of galaxies and are thus intimately connected to the growth rate of cosmic 
structure \cite{Kaiser:1987qv, 1998ASSL..231..185H}. Over the last 10 years, as the size of spectroscopic surveys has 
increased, this effect has been exploited, allowing testable predictions of 
general relativity on large scales~\cite{Linder:2007nu,Zhang:2007nk,Jain:2007yk,Wang:2007ht,Guzzo:2008ac,Song:2008qt,Daniel:2010yt,Song:2010rm,Song:2010fg,Reyes:2010tr,Shafieloo:2012ms,Reid:2012sw,Yoo:2012vm}.
  
Geometric distortions are induced by distances along and perpendicular to 
the line of sight being fundamentally different. 
Measuring the ratio of galaxy clustering radially to transversely 
provides a probe of this, called the Alcock-Paczynski effect~\cite{Alcock:1979mp,Padmanabhan:2008ag,Gaztanaga:2008xz}. 
Assuming the incorrect cosmological model 
for the coordinate transformation from redshift space to comoving Cartesian 
space leaves a residual geometric distortion. 
In observations the geometric effect is convolved with RSD~\cite{1996MNRAS.282..877B, 1996ApJ...470L...1M}, but the fixed physical 
scale of baryon acoustic oscillations (BAO) can alleviate this covariance~\cite{2003ApJ...594..665B, 2005ApJ...633..560E}. 

The quantity and quality of data now allows the distinction of these effects 
through a full two dimensional (transverse-radial) analysis, rather than 
relying on a spherical average, a squashing (AP) ratio, or 
the lowest few multipoles of the clustering distribution.  Following the 
method tested in~\cite{Song:2013ejh}, we fit the clustering correlation function 
in the transverse-radial plane, paying particular attention to the baryon 
acoustic oscillation (BAO) ``ring of power''~\cite{Hu:2003ti} but using the 
full 2D clustering information. 

Another advantage of this analysis is that it is carried out in a substantially 
model independent manner, without assuming LCDM or any other 
dark energy model.  Instead we directly fit for the angular distance $D_A$, 
Hubble parameter $H$, and growth rate $\gth\sim d\delta/d\ln a$ 
simultaneously from the 2D clustering statistics. 
Variation of each of these give distinct distortions of the clustering power 
isocontours, 
including the BAO ring of power. We analyze the SDSS DR9 galaxies in the BOSS 
CMASS \cite{Ahn:2012fh} sample at an effective redshift of $\zeff=0.57$. 

This data has already given rise to several significant results in measuring 
cosmological distances, the first BAO detection in DR9 coming from 
\cite{Anderson:2013oza} and \cite{2013A&A...552A..96B}. This 
was followed by a more detailed study which found the distance ratio 
$D_V(z=0.57)=2094\pm34$ Mpc 
\cite{Anderson:2013oza} using typical correlation functions and power 
spectra analyses, where $D_V$ is a spherically averaged distance measure. 
Anisotropic analysis using ``clustering wedges'' placed 
tight constraints on the angular diameter distance and the Hubble constant: 
$D_A(z=0.57)/r_s(z_d)=9.03 \pm 0.21$ and $cz/(r_s(z_d)H(z)) = 12.14 \pm 0.43$ 
\cite{Sanchez:2013uxa,Kazin:2013rxa}. These measurements were confirmed by 
\cite{Reid:2012sw} and \cite{2013MNRAS.429.1514S} used the full shape of the 
monopole and quadrupole correlation functions to obtain estimates of $H(z)$, $D_A$ 
and the growth rate $d\delta/d\ln a$. This reduced set of cosmological observables was then used to place tight 
constraints on the cosmological parameters \cite{2013MNRAS.429.1514S, 2013MNRAS.428.1116R}. 

Angularly-averaged statistics, such as the multipoles mentioned above, successfully placed constraints on cosmological parameters. However, such statistics become complicated when one considers excluding data along the line-of-sight, 
e.g.\ that are much noisier than the data perpendicular to it \cite{Scoccimarro:2004tg} or are difficult to model accurately because of velocity or nonlinear 
effects. It is thus meaningful to present the analysis of the correlation function in the transverse-radial plane, without angle averaging, as a complementary 
method. 

Another advantage of using the full 2D correlation function is that one can 
easily distinguish between the geometric and velocity (RSD) effects, 
clarifying the physical interpretation. The 2D correlation function including the BAO scale was first analyzed by \cite{2008ApJ...676..889O} but the analysis relied on linear theory \cite{2004ApJ...615..573M}. 
In \cite{Song:2013ejh} we developed a formalism that predicts the correlation function in the 2D plane with nonlinear perturbation theory. 
Following the method tested in~\cite{Song:2013ejh}, we here fit the 
clustering correlation function in the transverse-radial plane to data. 

The outline of this paper proceeds as follows. In Sec.~\ref{sec:theory} we 
briefly review the analysis method and treatment of 
nonlinearities and velocity effects.  Section~\ref{sec:measure} details 
the measurement procedure including estimation of the covariance matrix. 
The results are presented in Sec.~\ref{sec:results} and the implications for 
cosmological models are discussed in Sec.~\ref{sec:cosmology}. We conclude in 
Sec.~\ref{sec:concl}, with Appendix~\ref{sec:zeff} exploring cautions 
regarding interpretation of $\zeff$ at 
the accuracy of next generation surveys.

\section{Theoretical Model} \label{sec:theory} 

The two--point correlation function of galaxy clustering, $\xi_s$, 
is described as a function of $\sigma$ and $\pi$ in the distant-observer 
limit, where $\sigma$ and $\pi$ are the transverse and the radial directions 
with respect to the observer. 
As mentioned in the Introduction, several effects give rise to anisotropy 
between these directions. In the linear density perturbation regime, RSD 
causes the clustering pattern to be squeezed along the line of sight 
(i.e.\ the $\pi$-direction), leading to an apparent enhancement of the 
amplitude of the observed correlation function. This is known as Kaiser 
effect~\cite{Kaiser:1987qv}. On the other hand, in the non--linear regime, 
the random virial motions of galaxies residing in halos introduce elongated 
clustering along the line of sight, which is dubbed the Finger of God effect 
(FoG). This dispersion effect has significant impact, and even on large 
scales (in linear theory), a simple description of $\xi_s(\sigma,\pi)$ 
using the Kaiser formula alone may not be adequate along the $\pi$ direction 
(e.g., \cite{Scoccimarro:2004tg}). In our previous paper, we combined this 
dispersion effect with the Kaiser formula to analyze two--dimensional 
anisotropy structure of DR7 catalogue~\cite{Song:2010kq}. 

The precision of the updated DR9 clustering catalog is greatly improved. 
Due to this improvement, systematic uncertainties in accounting for the 
anisotropic clustering effects gain greater influence. Therefore we here 
employ improved distortion models to analyze the better precision maps. 
Due to a strong correlation between density and velocity fields, the mapping 
between real and redshift space is intrinsically non--linear~\cite{Taruya:2010mx}. In general, it appears as a not--closed iteration of polynomials for which a more elaborate description than simple factorized formulation needs to be 
used. However at large separation several leading polynomials dominate. In addition, we apply the non--linear correction terms using the resummed perturbation theory called {\tt RegPT}~\cite{Taruya:2007xy,Taruya:2012ut}. When restricting 
analysis to the quasi-linear regime, the result is the non--linear portions 
of the power spectra are better separated from the linear spectra, for which the assumption of perfect cross--correlation between density and velocity fields  is verified. 

In brief, we adopt the redshift-space power spectrum, $\tilde{P}(k,\mu)$, given in Ref.~\cite{Taruya:2010mx}, which can be recast as 
\ba
\label{eq:pkred_in_Q}
\tilde{P}(k,\mu) =\sum_{n=0}^4\,Q_{2n}(k)\mu^{2n}\,G^{\rm FoG}(k\mu\sigma_p)\,,
\ea
where $\sigma_p$ is a free parameter representing small scale velocity 
effects. Our previous analysis suggests that as long as we consider the weakly nonlinear scales, cosmological analysis can be made independently of the functional form of FoG effect. The functions $Q_{2n}$ are given in~\cite{Song:2013ak}.

From the power spectrum one can compute the correlation function by Fourier 
transform. The redshift-space correlation function $\xi(\sigma,\mu)$ is generally expanded as
\ba\label{eq:xi_eq}
\xi^{s}(\sigma,\pi)&=&\int \frac{d^3k}{(2\pi)^3} \tilde{P}(k,\mu)e^{i{\bf k}\cdot{\bf s}}\nn\\
&=&\sum_{\ell\ {\rm even}}\xi_\ell(s) {\cal P}_\ell(\nu)\,,
\ea
with ${\cal P}$ being the Legendre polynomials. Here, we define $\nu=\pi/s$ and $s=(\sigma^2+\pi^2)^{1/2}$. The moments of correlation function are given in~\cite{Song:2013ejh}. Here we include the moments up to ${\ell}=6$, since the 
higher-order moments $\ell\ge 8$ are shown to contribute negligibly.

\section{Measurements} \label{sec:measure} 

\begin{figure*}
\begin{center}
\resizebox{3.2in}{!}{\includegraphics{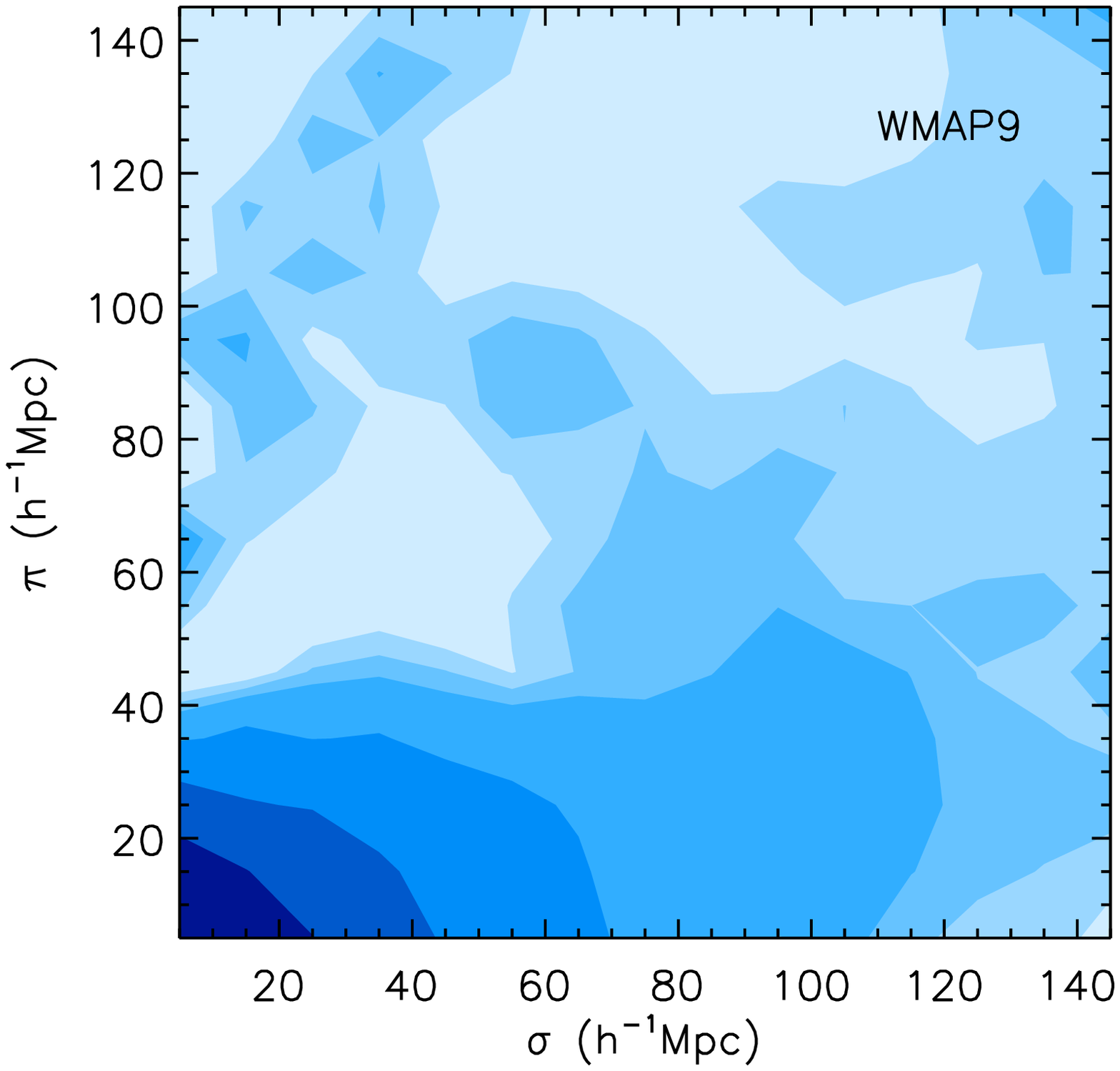}}\hfill 
\resizebox{3.2in}{!}{\includegraphics{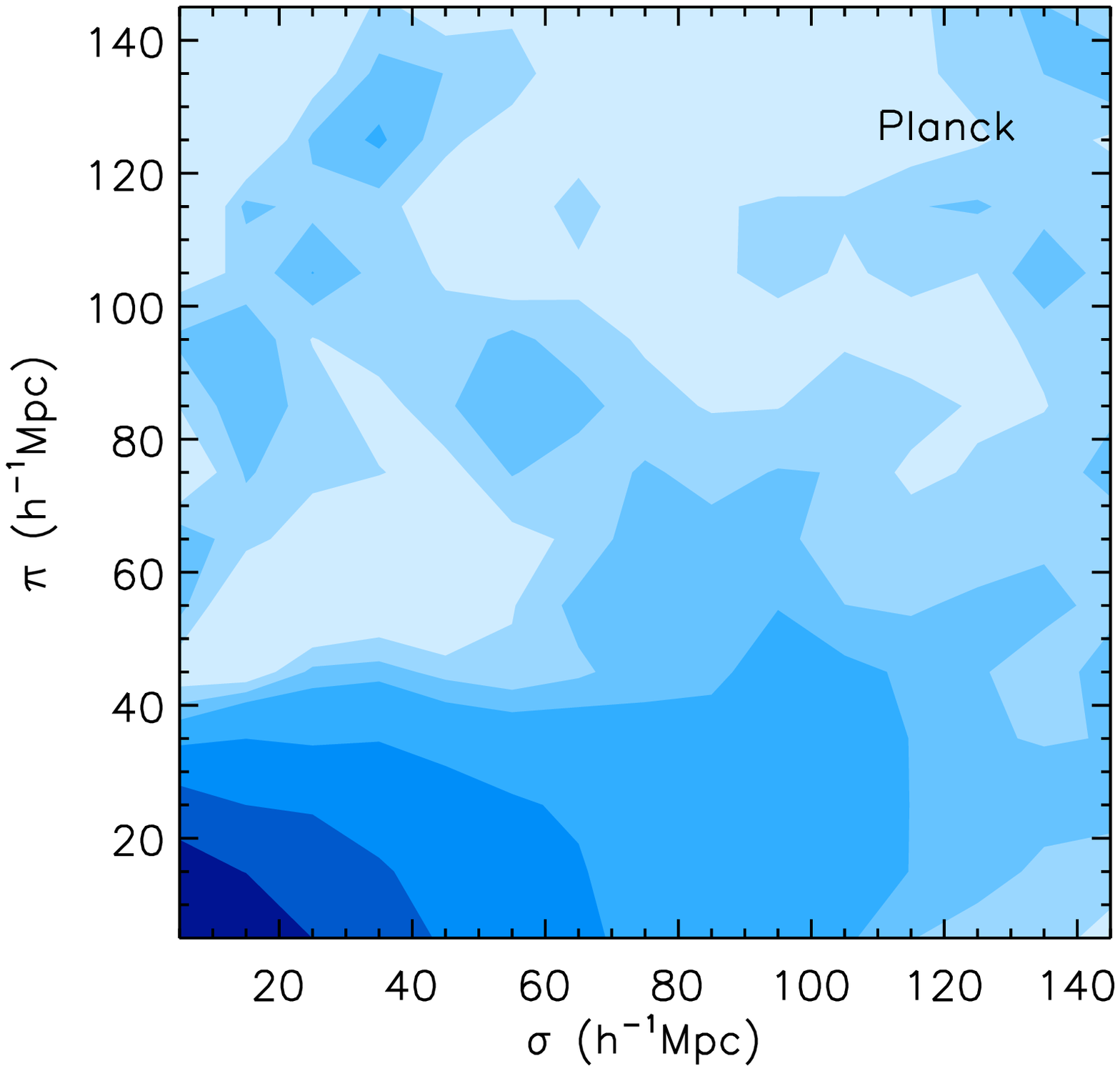}}
\end{center}
\caption{The measured 2D clustering correlation function $\xi(\sigma,\pi)$ 
is plotted, adopting early universe priors from WMAP9 (left) or Planck 
(right).} 
\label{fig:measured_xi}
\end{figure*}

\subsection{Data Sample} 

We use data from the Sloan Digital Sky Survey \citep[SDSS; ][]{2000AJ....120.1579Y}, Data Release 9 (DR9). SDSS has mapped over one quarter of the sky in five photometric bands down to a limiting magnitude of $r\sim 22.5$. The photometric data is reduced and from it are selected targets for followup spectroscopy. 
The spectroscopic survey, known as the Baryon Oscillation Spectroscopic Survey (BOSS), is designed to obtain spectra for $\sim1.5$ million galaxies over a 10,000 square degree footprint. 

In an effort to control the evolution of galaxy bias over large redshift ranges the BOSS targets are selected in such a way as to have approximately constant stellar mass (CMASS). This is obtained using colour selections based on the passive galaxy template of \citep{2009MNRAS.394L.107M}. The majority of CMASS galaxies are bright, central galaxies (in the halo model framework) and are thus highly biased ($b\sim2$) \citep{2013MNRAS.432..743N}.

The CMASS sample~\cite{CMASS} 
is defined by 
\ba
z&>&0.4\\
17.5<i_{cmod}&<&19.9\notag\\
r_{mod}-i_{mod}&<&2.0\notag\\
d_{\perp}&>&0.55\notag\\
i_{fib2}&<&21.5\notag\\
i_{cmod}&<&19.86+1.6(d_{\perp}-0.8)\notag\\
i_{psf}-i_{mod}&>&0.2+0.2(20.0-i_{mod})\notag\\
z_{psf}-z_{mod}&>&9.125-0.46z_{mod}\ ,\nn 
\ea 
where the last two conditions provide a star-galaxy separator and $d_{\perp}$ is defined as \citep{2006MNRAS.372..425C},
\ba 
d_{\perp}=r_{mod}-i_{mod}-(g_{mod}-r_{mod})/8.0\ . 
\ea

Each spectroscopically observed galaxy is weighted to account for three 
distinct observational effects: redshift failure, $w_{fail}$; minimum 
variance, $w_{FKP}$; and angular variation, $w_{sys}$. These weights are 
described in more detail in \cite{2012MNRAS.427.3435A} and 
\cite{2012MNRAS.424..564R}. Firstly, galaxies that lack a redshift due to 
fiber collisions or inadequate spectral information are accounted for by 
reweighting the nearest galaxy by a weight $w_{fail}=(1+N)$, where $N$ is 
the number of close neighbours without an estimated redshift. Secondly, the 
finite sampling of the density field leads to use of the minimum variance FKP 
prescription \cite{1994ApJ...426...23F} where each galaxy is assigned a weight 
according to 
\begin{equation}
w^i_{FKP}=\frac{1}{1+n_i(z)P_0}\ , 
\end{equation}
where $n_i(z)$ is the comoving number density of galaxy population $i$ 
at redshift $z$ and 
one conventionally evaluates the weight at a constant power 
$P_0 \sim P(k=0.1\,h/{\rm Mpc}) \sim 2\times 10^{4}\,h^{-3}\,{\rm Mpc^3}$, 
as in \cite{2012MNRAS.427.3435A}. (But see Appendix~\ref{sec:zeff}.) 

The third weight corrects for angular variations in completeness and 
systematics related to the angular variations in stellar density that make 
detection of galaxies harder in over-crowded regions of the sky 
\cite{2012MNRAS.424..564R}. The total weight for each galaxy is then the 
product of these three weights, $w_{total}=w_{fail} w_{FKP} w_{sys}$.  
The random catalog points are also weighted but they only include the 
minimum variance FKP weight. 

The CMASS galaxy sample is distributed over the range $0.43<z<0.7$, with an 
effective redshift 
\ba
\zeff=\frac{\sum^{N_{gal}}_iw_{FKP,i}\,z_i}{\sum^{N_{gal}}_iw_{FKP,i}}\ , 
\ea 
giving the value $\zeff=0.57$. 
The effective volume 
\be
V_{\rm eff}=\sum\left(\frac{{n}(z_i)P_0}{1+{n}(z_i)P_0}\right)^2\Delta V(z_i) 
\ , 
\ee
where $\Delta V(z)$ is the volume of a shell at redshift $z$, 
is $V_{\rm eff}\sim 2.2\,$Gpc$^3$. 

\subsection{Measuring the correlation function}\label{sec:2pcf_boss}

We compute the redshift-space 2-dimensional correlation function 
$\xi(\sigma,\pi)$ using the BOSS DR9 galaxy catalog 
\cite{2012MNRAS.427.3435A}. We perform the coordinate transforms for two 
fiducial spatially-flat cosmological models: WMAP9 ($\omega_b=0.02264$, $\omega_c=0.1138$, $h=0.70$) and PLANCK ($\omega_b=0.022068$, $\omega_c=0.12029$, $h=0.67$). Although the parameter fitting procedure should be insensitive to the choice of fiducial model, we perform this check for consistency. 

We estimate the correlation function using the standard Landy-Szalay 
estimator~\cite{1993ApJ...412...64L},
\begin{equation}
\xi(\sigma,\pi)=\frac{DD-2DR+RR}{RR}\ ,
\end{equation}
where $DD$ is the number of galaxy--galaxy pairs, $DR$ the number of galaxy-random pairs, and $RR$ is the number of random--random pairs, all separated by a distance
$\sigma\pm\Delta\sigma$ and $\pi\pm\Delta\pi$. Each pair is weighted by the product of the individual weightings of each point.

The random point catalogue constitutes an unclustered but observationally representative sample of the BOSS CMASS survey. The points are chosen to reside within the survey geometry and the redshifts are obtained via the random shuffle method of \cite{2012MNRAS.424..564R}. The randoms are also assigned completeness weights, just as for the galaxies. To reduce the statistical variance of the estimator we use $\sim20$ times as many randoms as we have galaxies. 

We calculate the correlation function in 15 bins of dimension $10\mpcoh$, 
linearly spaced in the range $0<\sigma,\pi<150\mpcoh$.  The resulting two 
point correlation function in Fig.~\ref{fig:measured_xi} shows the typical 
Kaiser \cite{Kaiser:1987qv} compression at small $\sigma$ (near to the line of sight) 
and the emergence of the 2D BAO ring at 
$\sqrt{\sigma^2 + \pi^2}\approx 100~\mpcoh$.

\subsection{Covariance matrix} 

In addition to the correlation function we need to know the errors on it. 
Because different bins of the correlation function are strongly correlated, 
it is necessary to estimate a covariance matrix to give correct constraints 
on cosmological parameters. 
As in our previous paper \cite{Song:2013ejh}, we use the mock galaxy catalog created by \cite{Manera:2012sc}. 
This catalog has the same survey geometry and number density as the CMASS galaxy sample that was used in our analysis and 611 mock realizations were created 
using second-order Lagrangian perturbation theory (2LPT) for the galaxy 
clustering. 

For each realization we compute the correlation function as we did for the observed catalog in section \ref{sec:2pcf_boss} and obtain a covariance matrix by 
\ba 
{\rm Cov}&&(\xi_i,\xi_j)=\\ 
&&\frac{1}{N_{mocks}-1}\sum^{N_{mocks}}_{k=1}[\xi_k({\bf r}_i)-\overline{\xi}({\bf r}_i)][\xi_k({\bf r}_j)-\overline{\xi}({\bf r}_j)]\ ,\nn 
\ea 
where $N_{mocks}=611$, $\xi_k({\bf r}_i)$ represents the value of the 
correlation function in the $i$th bin of $(\sigma,\pi)$ in the $k$th 
realization, and $\overline{\xi}({\bf r}_i)$ is the mean value of 
$\xi_k({\bf r}_i)$ over all the realizations. 
We can then obtain the normalized covariance matrix as 
\begin{equation}
\hat{C}_{ij}=\frac{{\rm Cov}(\xi_i,\xi_j)}{\sqrt{{\rm Cov}(\xi_i,\xi_i){\rm Cov}(\xi_j,\xi_j)}}\ .
\end{equation} 
In order to reduce the statistical noise in our covariance matrix, we perform a singular value decomposition (SVD)
of the matrix as done in \cite{Song:2010kq, 2010JCAP...01..025S},
\begin{equation}
\hat{C}_{ij} =  U_{ik}^{\dag} D_{kl} V_{lj},
 \end{equation}
 where $U$ and $V$ are orthogonal matrices that span the range and
the null space of ${\hat C}_{ij}$, and $D_{kl}=\lambda^2\delta_{kl}$, a diagonal
 matrix with the singular values along the diagonal. 

When using SVD the $\chi^2$ value becomes more difficult to interpret as it changes as 
one cuts the noisiest eigenvalues.  However we establish that the reduced $\chi^2$ converges 
to a constant value above 250 modes. To be conservative we use 350 out of 400 available modes. 

The estimate of the covariance matrix obtained from a finite number of realizations is 
necessarily biased (\cite{2007A&A...464..399H}, see also 
\cite{Kazin:2013rxa}). 
To obtain the unbiased covariance matrix $C$, we multiply the original covariance $\hat{C}$ by a correction factor  
\begin{equation}
C^{-1}=\frac{N_{mocks} - N_{bins}-2}{N_{mocks}-1}\ \hat{C}^{-1}\ ,
\end{equation}
where $N_{bins}$ is the number of bins of $\xi(\sigma,\pi)$ used for the 
analysis. 

\section{Results of 2D Anisotropy Analysis} \label{sec:results} 
\subsection{Fitting method}

\begin{figure*}
\begin{center}
\resizebox{3.2in}{!}{\includegraphics{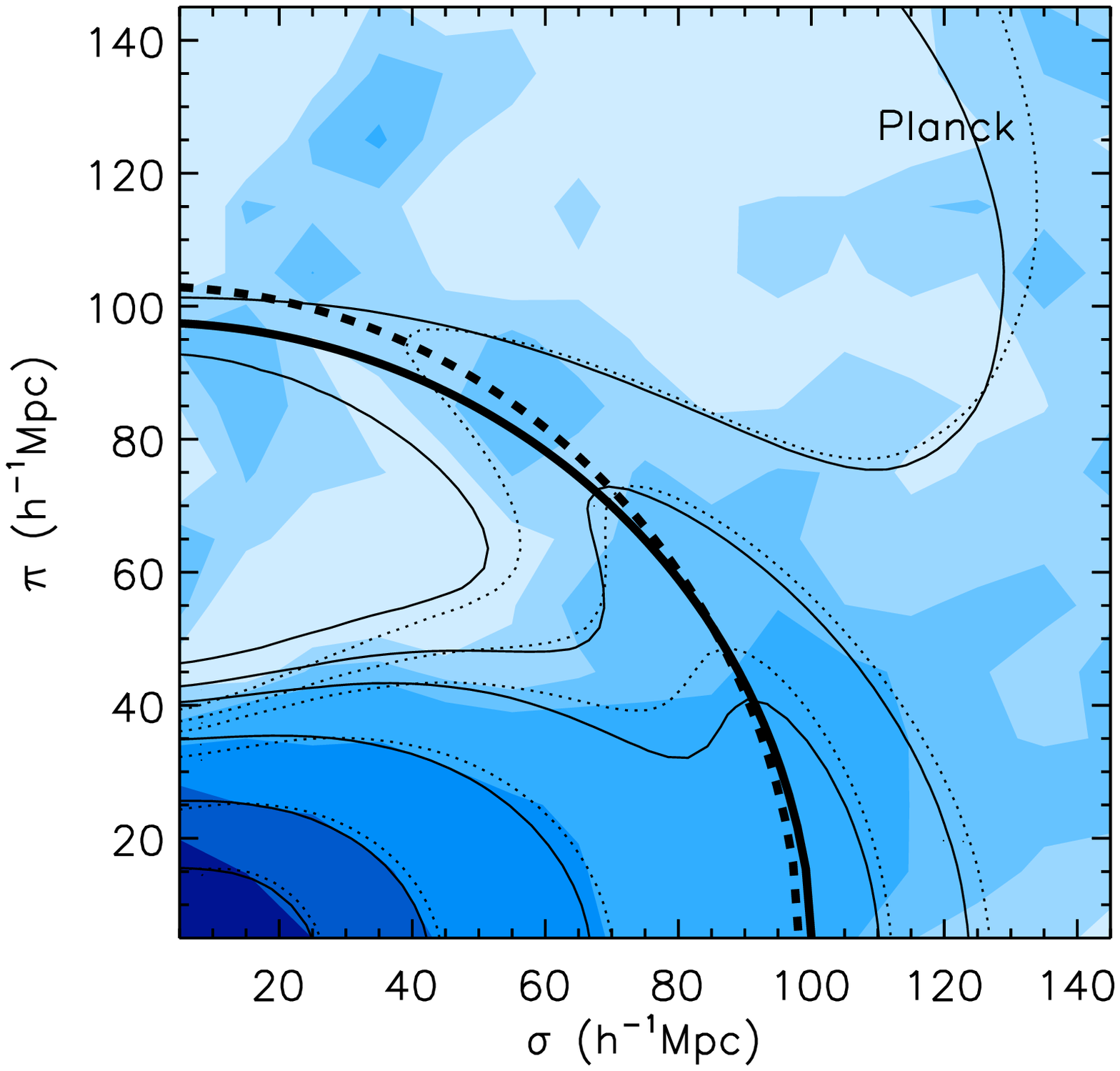}}\hfill
\resizebox{3.2in}{!}{\includegraphics{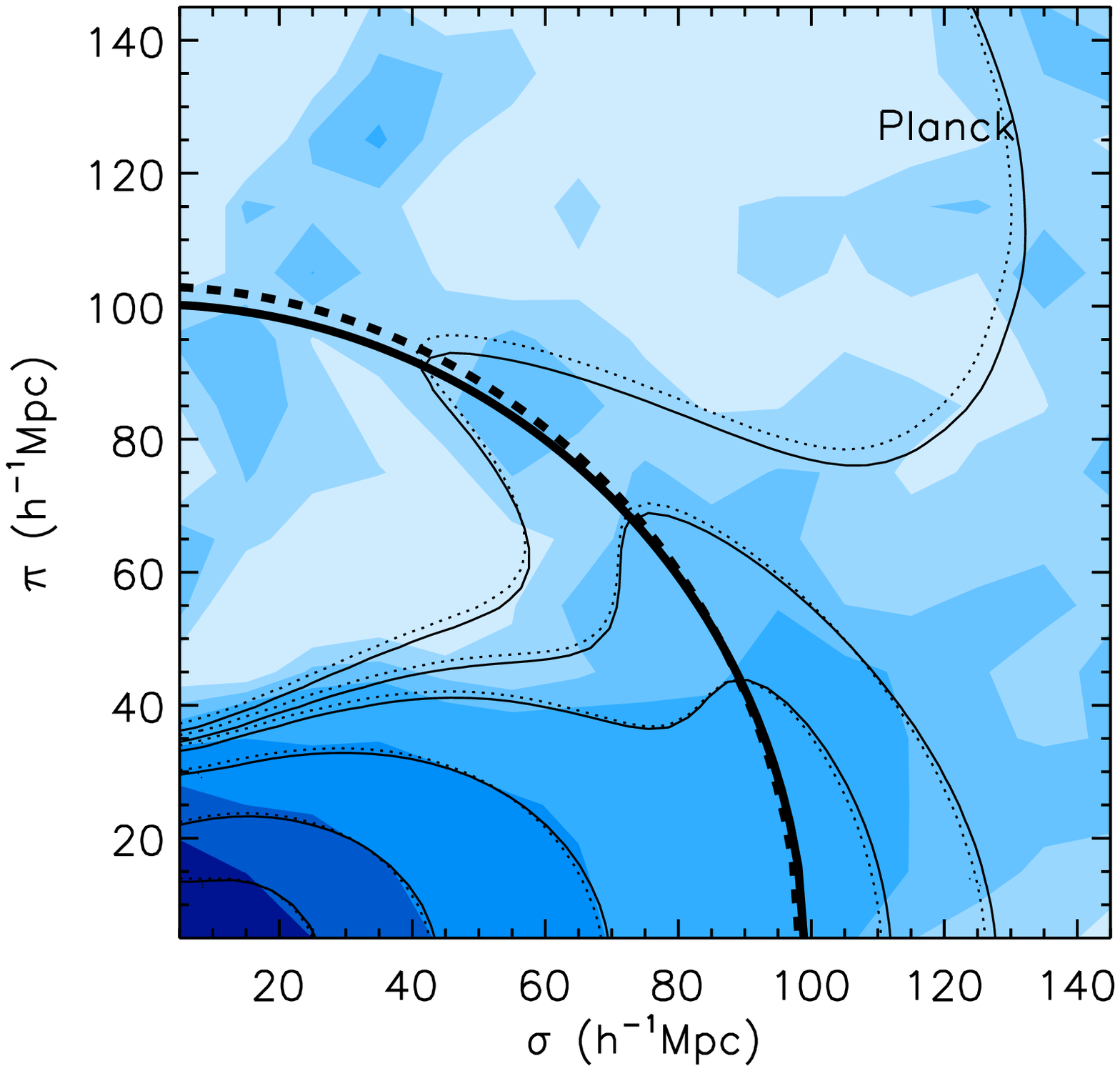}}
\end{center}
\caption{The measured, best fit, and LCDM-predicted versions of 
$\xi(\sigma,\pi)$ 
are plotted, using the Planck early universe prior. The blue filled contours 
represent the measured $\xi(\sigma,\pi)$, and the black unfilled contours 
represent the best fit $\xi(\sigma,\pi)$ with two different 
$\sigma_{\rm cut}$. The left panel uses $\sigma_{\rm cut}=20\mpcoh$ in the 
fit, and the right panel uses $\sigma_{\rm cut}=40\mpcoh$. 
The quarter rings denote the 2D BAO ring, for the fit (solid) and 
Planck LCDM prediction (dotted).} 
\label{fig:varyingcut_Planck}
\end{figure*}

To fit the correlation function with as model independent cosmology 
inputs as possible, we assume that the shape of the power spectra is 
given by the early universe conditions measured by CMB experiments. 
We denote this primordial spectrum convolved with the transfer function 
as $D_m(k)$. This 
then evolves coherently through all scales from the last scattering 
surface. That is, the growth occurs through a time-varying, scale-independent 
amplitude growth factor (this assumption breaks down in theories that 
introduce significant scale dependence, such as some modified gravity 
theories). Propagating this through to cosmological parameters 
requires assumptions on the cosmological model, e.g.\ the nature of dark 
energy. To remain substantially model independent we use the growth rate 
itself as our variable. 

The power spectra are then given by 
\ba\label{eq:PkDmG}
P_{bb}(k,a)&=&D_{m}(k)G_{b}^2(a),\nn\\
P_{\Theta\Theta}(k,a)&=&D_m(k)G_{\Theta}^2(a)\,,
\ea
where $G_{b}$ and $G_{\Theta}$ denote the growth functions of density and peculiar velocity. We define here $G_{b}=b\,G_{\delta_m}$ where $b$ is the standard linear bias parameter between galaxy tracers and the underlying dark matter density. The expression of $D_m(k)$ is available in~\cite{Song:2010kq}, and assumed to be given precisely by CMB experiments, such as WMAP9 and Planck 
experiments. We refer to this as an early universe prior. 
We incorporate the uncertainty from the CMB anisotropy data in the amplitude 
determination of the initial spectra, $A_S^2$, into the growth function $G_X$, 
i.e.\ $G_X=G^*_XA_S/A_S^{*}$ where $G^*_X$ is the intrinsic growth function. 

The clustering correlation function $\xi(\sigma,\pi)$ is measured in comoving 
distances, while galaxy locations use angular coordinates and redshift in 
galaxy redshift surveys. A fiducial cosmology is required for conversion 
into comoving space. We use the best fit LCDM universe to WMAP9 or Planck. 
The observed anisotropy correlation function using this model is transformed 
into true comoving coordinates using the transverse and radial distances, 
involving $D_A$ and $H^{-1}$, respectively. The approximate fiducial maps 
are created by rescaling the transverse and radial distances, using 
\ba
\sigma^{\rm fid}&=&\frac{D_A^{\rm fid}}{D_A^{\rm true}}\ \sigma^{\rm true} \nn \\
\pi^{\rm fid}&=&\frac{H^{-1\,\rm fid}}{H^{-1\,\rm true}}\ \pi^{\rm true} \ , 
\ea 
where ``fid'' and ``true'' denote the fiducial and true distances. 
Thus the theoretical $\xi$ with potentially true $D_A$ and $H^{-1}$ is 
fitted to the observed $\xi^{\rm fid}$ using the rescaling in the above 
equations.

\begin{figure*}
\begin{center}
\resizebox{3.2in}{!}{\includegraphics {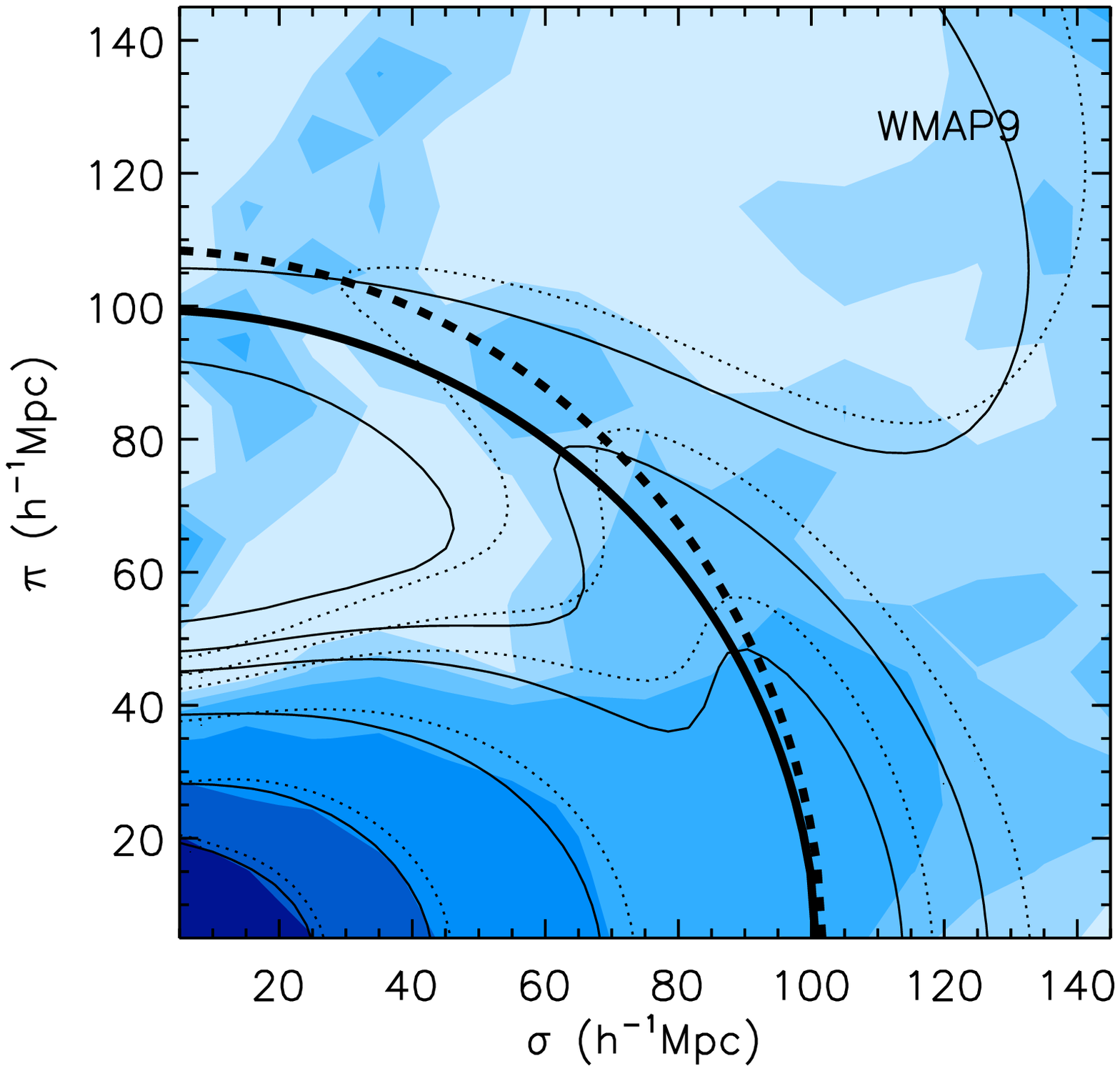}}\hfill
\resizebox{3.2in}{!}{\includegraphics{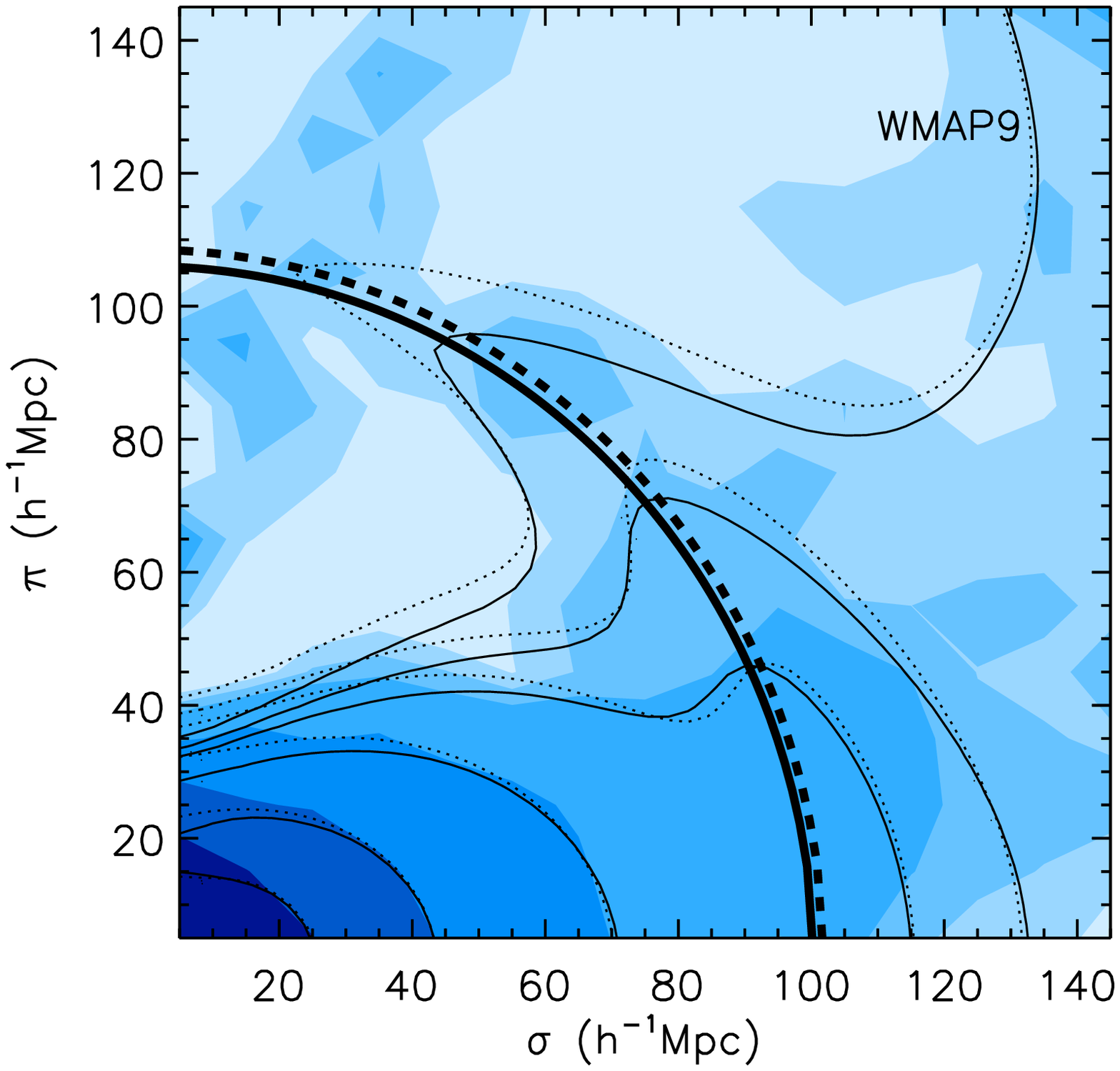}}
\end{center}
\caption{As Fig.~\ref{fig:varyingcut_Planck}, but using WMAP9 instead of 
Planck.} 
\label{fig:varyingcut_WMAP9}
\end{figure*}

Given the early universe prior on the power spectrum shape, 
both distances and growth functions are measured simultaneously to high precision~\cite{Song:2012gh}. This holds even without assuming the FRW integral 
relation between $H^{-1}$ and $D_A$. Thus we do not have to assume any 
particular cosmological model or restrict to zero curvature or LCDM. 

Finally, we introduce a parameter representing non--linear contamination 
to the power spectra of the density and velocity fields. Even on linear 
scales the damping effect on the power spectrum amplitude caused by random 
galaxy motions still remains. This is described by the Gaussian model for 
the FoG effect in Eq.~\ref{eq:pkred_in_Q}, with $\sigma_p$ a free parameter 
giving the velocity dispersion. However, the non--perturbative damping 
effects are not fully understood, and the Gaussian model may be insuficient 
on non--linear scales. We therefore do not use the measured $\xi(\sigma,\pi)$ 
for bins in which this breakdown is likely. Two cut--off's are used: 
1) $s_{\rm cut}$ represents the scales on which non--linear description of 
$\Delta P_{XY}$ is uncertain, and 2) $\sigma_{\rm cut}$  represents the 
scales on which Gaussian FoG functional form may not be appropriate. These 
are set to be $s_{\rm cut}=50\mpcoh$ and $\sigma_{\rm cut}=40\mpcoh$ 
(although we also consider $\sigma_{\rm cut}=20\mpcoh$). This strategy was 
tested and proved valid using simulations in our previous work. We follow 
the same method as presented in Song, Okumura and Taruya (2013)~\cite{Song:2013ejh}.

In summary, we have $G_b$ and $G_{\Theta}$ to describe growth functions, 
$D_A$ and $H^{-1}$ to fit distance measures, and $\sigma_p$ to model the 
FoG effect. The form of the FoG is taken to be Gaussian and the shape of 
the linear spectra is assumed to be given as an early universe prior by 
CMB experiments.

\subsection{Cut--off scales and 2D BAO ring}

\begin{figure*}
\begin{center}
\resizebox{3.5in}{!}{\includegraphics {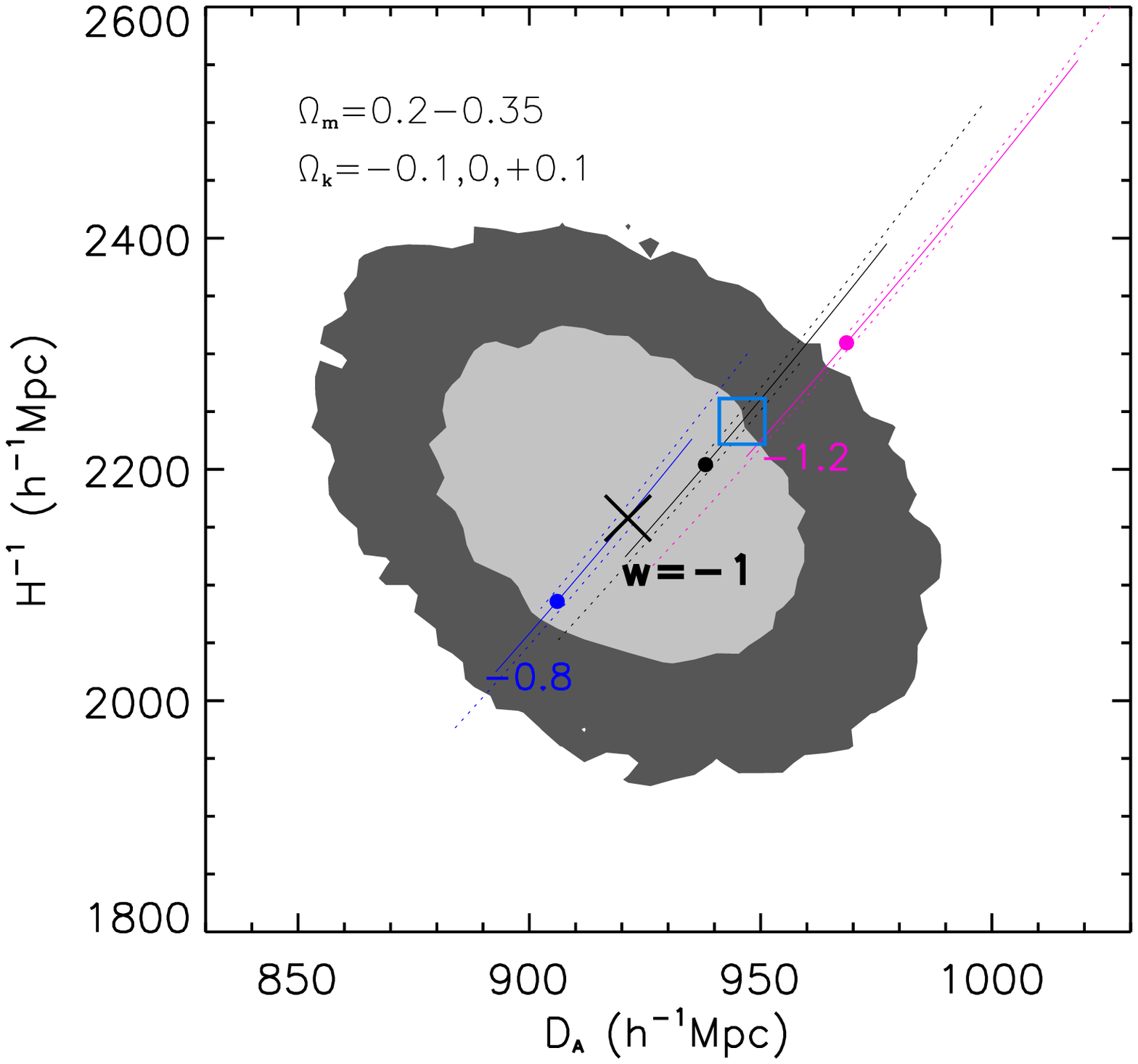}}
\resizebox{3.5in}{!}{\includegraphics {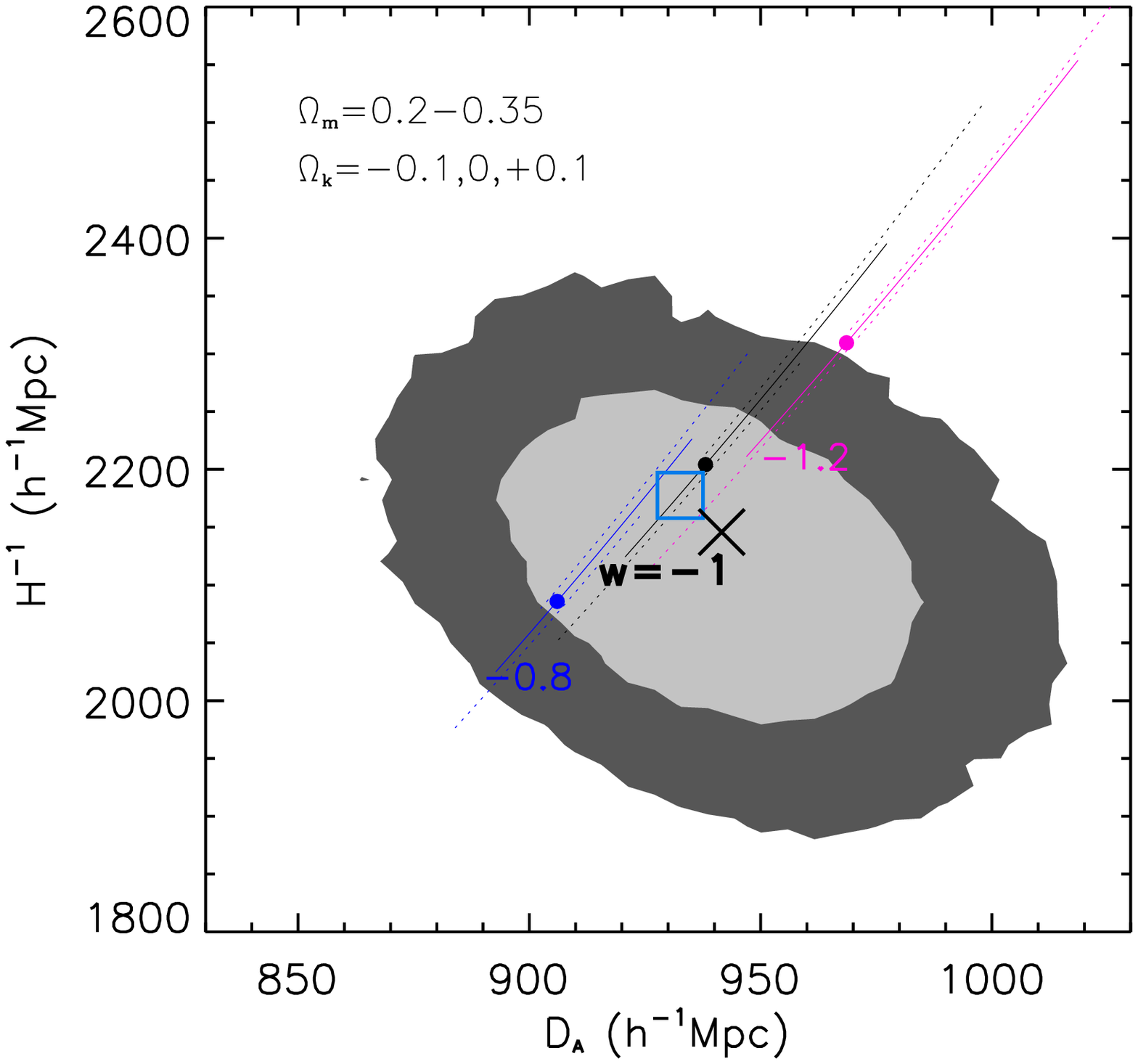}}
\end{center}
\caption{The 68\% and 95\% confidence contours from the galaxy clustering 
data are plotted in the $D_A-H^{-1}$ plane, with the best fit denoted by the 
large X.  The values predicted from the CMB within LCDM cosmology are 
shown by the blue square. The left panel uses the WMAP9 early universe prior 
on the power spectrum shape, while the right panel uses the Planck prior. 
Overlaid are theory curves giving the relation between the two cosmological quantities 
within certain cosmologies; note that all standard cosmologies lie in a 
restricted band.  In addition to LCDM (black solid), we show wCDM with 
$w=-0.8$ (blue) or $w=-1.2$ (magenta), and their generalizations to include 
spatial curvature (dotted).  Each curve covers the range of $\om=0.2$ at their 
upper ends to $\om=0.35$ at their lower ends, with large dots showing the 
$\om=0.3$ case.  
} 
\label{fig:dah}
\end{figure*}

First, we investigate the appropriate cut--off scales. The $s_{\rm cut}$ is introduced due to the uncertainty of the resummed perturbation theory {\tt RegPT} at smaller scales. It is conservatively set to be $s_{\rm cut}=50\mpcoh$ which allows the perfect cross--correlation between density and velocity fields. 
In addition $\sigma_{\rm cut}$ is used because the improved $\xi(\sigma,\pi)$ in Eq.~\ref{eq:xi_eq} is not applicable at bins in which the higher order terms of non--perturbative effect are dominant along the line of sight. It was set to be $\sigma_{\rm cut}=20\mpcoh$ in~\cite{Song:2013ejh}, and reproduced the true values successfully. But we find that it may be too ambitious for the actual 
DR9 CMASS catalogue. 

When the broadband shape of spectra and the distance measures are known, the 
2D BAO ring is invariant to the changes of the coherent galaxy bias and 
coherent motion growth function. When $G_b$ increases/decreases, the BAO tip 
points coherently move counter--clockwise/clockwise. When $G_{\Theta}$ 
increases/decreases, the BAO tip points move toward/away from the pivot 
point (equal radial and transverse separation). If the correct distance model 
is known, the tip points of BAO peaks form 
an invariant ring regardless of galaxy bias and coherent motion. The ratio 
between the observed transverse and radial distances varies with the assumed 
cosmology and, if the shape of an object is {\it a priori} known, can provide 
a measure of $HD_A$ (AP test). 

The outer measured $\xi(\sigma,\pi)$ contours are too vague to reveal 
detailed BAO peak structure, but those peak points can define the measured 
2D BAO ring. Figure~\ref{fig:varyingcut_Planck} shows the 2D correlation 
function contours, and the best fit 2D BAO rings. The left and right panels 
use $\sigma_{\rm cut}=20\mpcoh$ and $40\mpcoh$, respectively. If 
the correlation function model is accurate down to 
$\sigma_{\rm cut}=20\mpcoh$ then the two rings should be consistent. 
However, the 2D BAO ring using $\sigma_{\rm cut}=20\mpcoh$ not only 
disagrees with that using $\sigma_{\rm cut}=40\mpcoh$ but also from the 
measured circle. 

Basically the small, nonlinear scales where the model is imperfect are 
distorting the results at all scales. This can be seen by looking at 
several inner contours at small scales, those corresponding to 
$\xi=(0.2,0.06,0.016,0.005)$. In the left panel the solid curves attempt to 
fit tightly the small scale contours very close to the line of sight, at 
the price of a poor fit to the large scale, linear contours. By contrast, 
in the right panel with $\sigma_{\rm cut}=40\mpcoh$ the residual 
non--perturbative effects are observed clearly in the inner contours, but 
the linear contours are better behaved. This problem with an overambitious 
use of small scales is seen as well in Fig.~\ref{fig:varyingcut_WMAP9} 
using the WMAP9 early universe prior instead. 

Therefore we use more conservative bound at $\sigma_{\rm cut}=40\mpcoh$. 
We tested our final results using different $\sigma_{\rm cut}$ at 20, 30, 40, 
and 50 $\mpcoh$ and found they converged for $\sigma_{\rm cut}\ge40\mpcoh$. 
The effect on cosmology of using a cut allowing more of the non--linear 
regime is discussed in Sec.~\ref{sec:cosmology}. 

The dashed contours in Fig.~\ref{fig:varyingcut_Planck} represent the 
$\xi(\sigma,\pi)$ of the Planck LCDM concordance model. They are derived using 
the fiducial $(D_A, H^{-1}, G_{\Theta})$, and best fit $(G_b,\sigma_p)$.  
The right panel of Fig.~\ref{fig:varyingcut_Planck} shows strong agreement 
between the derived best fit model and the theoretical Planck LCDM 
concordance model. 

For Fig.~\ref{fig:varyingcut_WMAP9} using the WMAP9 early universe prior, 
while the estimated 2D BAO ring agrees approximately with the measured 2D 
BAO ring, peak points along the ring do not well match to each other. The 
dashed contours here represent $\xi(\sigma,\pi)$ of the WMAP9 LCDM 
concordance model. Unlike the Planck case, the measured peak points shift 
toward the pivot point for the outer contour, less so for the inner 
contours. As discussed above, this is a signature of an increased velocity 
growth function; we expect the measured $\gth$ to be higher than fiducial 
in this case.

\subsection{The measured distances and growth functions}

We present the results for the measured distances and growth functions in 
Table~\ref{tab:measurements}. Our baseline value of 
$\sigma_{\rm cut}=40\mpcoh$ is used throughout this section.

\begin{table}
\begin{center}
\begin{tabular}{lccc}
\hline
\hline
Parameters & Fiducial values & Measurements \\
& With WMAP9 prior &  \\
\hline
$D_A\,(\mpcoh)$       & $946.0$ & $916.2^{+27.2}_{-25.4}$  \\
$H^{-1}\,(\mpcoh)$   & $2241.5$ & $2163.1^{+102.0}_{-85.8}$  \\
$G_b$   & $-$ & $1.07^{+0.07}_{-0.09}$ \\
$G_{\Theta}$ &$0.44$ & $0.51^{+0.09}_{-0.08}$ \\
$\sigma_p\,(\mpcoh)$ &$-$ & $1.0^{+4.6}$ \\
\hline
\hline
Parameters & Fiducial values & Measurements \\
& With Planck prior &  \\
\hline
$D_A\,(\mpcoh)$       & $932.6$ & $939.7^{+26.7}_{-32.6}$  \\
$H^{-1}\,(\mpcoh)$   & $2177.5$ & $2120.5^{+82.3}_{-100.6}$  \\
$G_b$   & $-$ & $1.11^{+0.07}_{-0.10}$ \\
$G_{\Theta}$ &$0.46$ & $0.47^{+0.10}_{-0.07}$ \\
$\sigma_p\,(\mpcoh)$ &$-$ & $1.2^{+4.0}$ \\
\hline
\hline
\end{tabular}
\end{center}
\caption{We summarize the values predicted by the CMB data and the values 
measured from the BOSS data of the distance quantities $D_A$ and $H^{-1}$ 
and the growth quantities $G_b$ and $G_{\Theta}$, as well as the velocity 
damping scale $\sigma_p$, with 68\% confidence level errors.  (The CMB data 
does not predict values of the astrophysical parameters $G_b$ and $\sigma_p$.)} 
\label{tab:measurements}
\end{table}

\begin{figure*}
\begin{center}
\resizebox{3.5in}{!}{\includegraphics {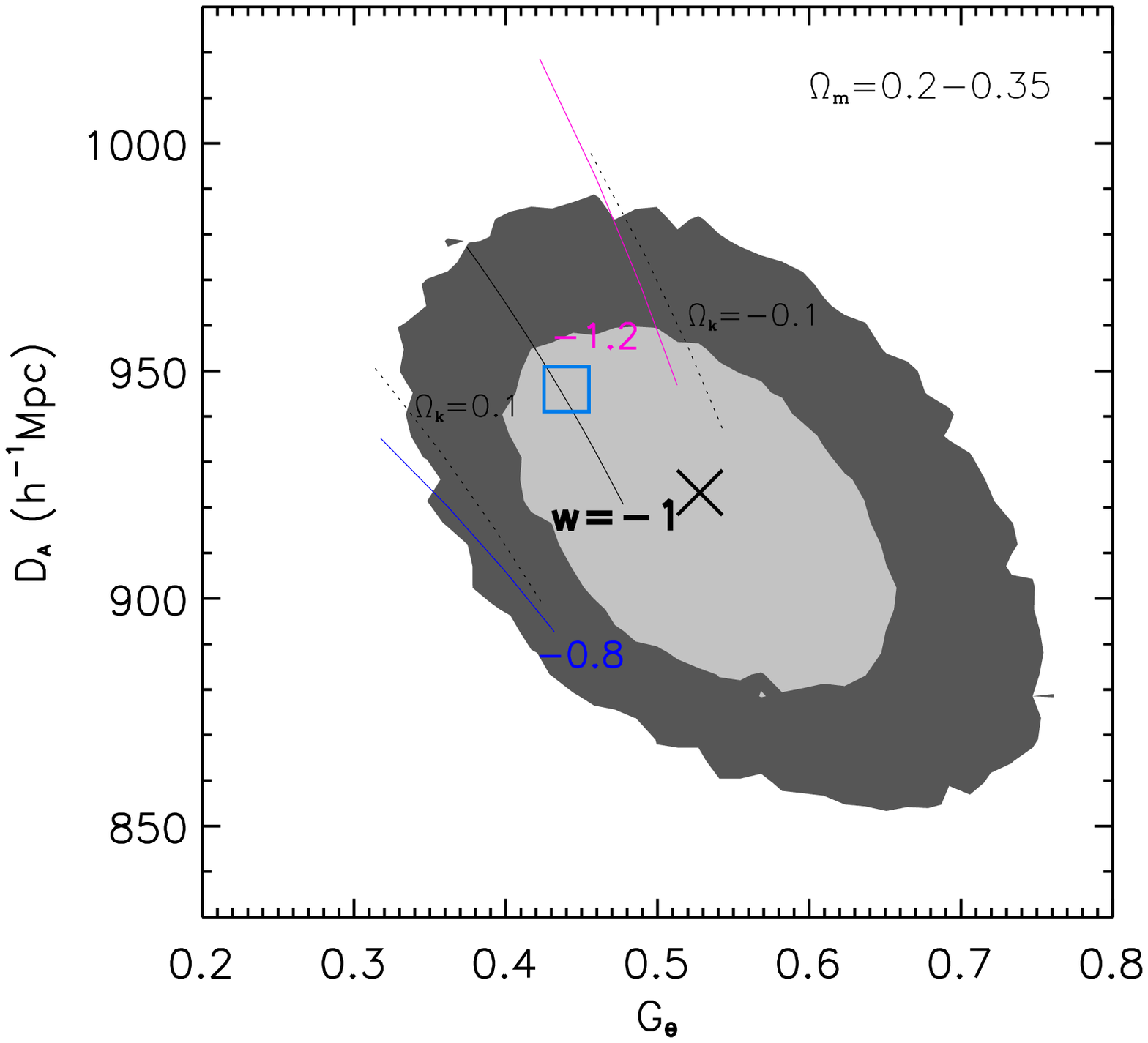}}
\resizebox{3.5in}{!}{\includegraphics {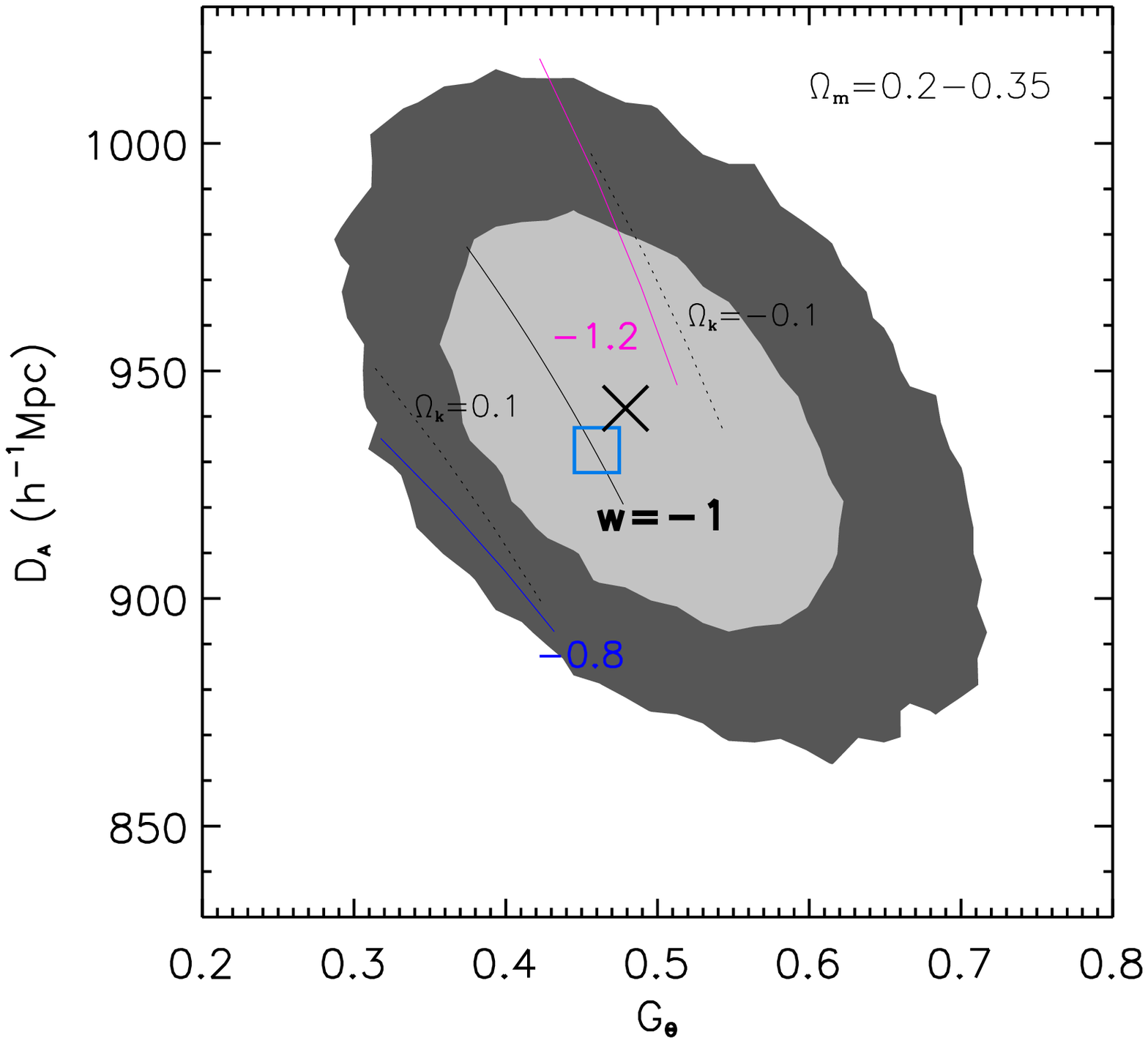}}
\end{center}
\caption{As Fig.~\ref{fig:dah} but for the $G_\theta-D_A$ plane.  Here 
the allowed cosmology band is wider (we do not plot the owCDM models). 
} 
\label{fig:dag}
\end{figure*}

The angular diameter distance $D_A$, related to transverse separations, 
is measured to be consistent with the LCDM predictions. Most uncertainties 
of anisotropic distortions are relevant to the radial direction, and it is 
expected that $D_A$ is not biased much. With the Planck early universe prior, 
$D_A$ is measured to be $939.7^{+26.7}_{-32.6}\mpcoh$, in excellent agreement 
with the Planck LCDM best fit prediction. Using the WMAP9 early universe prior, 
the measured $D_A$ is $1-\sigma$ from the WMAP9 LCDM prediction. 

The line of sight distance quantity $H^{-1}$, related to radial separations, 
is also measured to be consistent with LCDM predictions. For either the 
Planck or WMAP9 early universe priors the agreement is within $1-\sigma$. 
Greater tension is seen if one uses $\sigma_{\rm cut}=20\mpcoh$, which 
lowers the measured $H^{-1}$. 

As mentioned earlier, the growth functions influence the location of peaks 
along the rings of power (see \cite{Song:2013ejh} for illustrations). For 
the Planck early universe prior, the best fit peak structure is nearly 
identical to that predicted by the Planck LCDM model. The measured coherent 
growth function has $G_{\Theta}=0.47^{+0.10}_{-0.07}$, while the fiducial 
value is $0.46$. This measurement can be converted to a value at $\zeff=0.57$ 
of the standard parameter $f\sigma_8=0.48$, which is very close to the 
fiducial model value of 0.47. When the WMAP9 early universe prior is used, 
the measured $G_{\Theta}$ becomes bigger than LCDM prediction. Like the 
distance measurements, the measured $G_{\Theta}$ is offset by $\sim1-\sigma$. 

Note that $G_{\Theta}$ has a relatively large error, about $15-20\%$. 
This is partly caused by floating $\sigma_p$ as a free parameter. In the 
linear regime, when the first order contribution of the Gaussian FoG 
function dominates, this factor is nearly featureless and becomes 
significantly degenerate with coherent growth function. Using more non-linear 
scales (smaller $\sigma_{\rm cut}$) would break this degeneracy, reducing 
the error contour but introducing bias; we show this explicitly in 
Sec.~\ref{sec:cosmology}. 

The galaxy bias is estimated from the $G_b$ measurement. The bias $b$ is measured to be $1.9$ and $1.8$ for Planck and WMAP9 respectively. Those values are consistent with CMASS catalogues~\cite{Manera:2012sc}. The velocity dispersion $\sigma_p$ indicates the level of the FoG effect. For both Planck and WMAP9 cases, 
it is observed to be small, about $\sigma_p=1\mpcoh$, but with significant 
uncertainty.

\begin{figure*}
\begin{center}
\resizebox{3.5in}{!}{\includegraphics {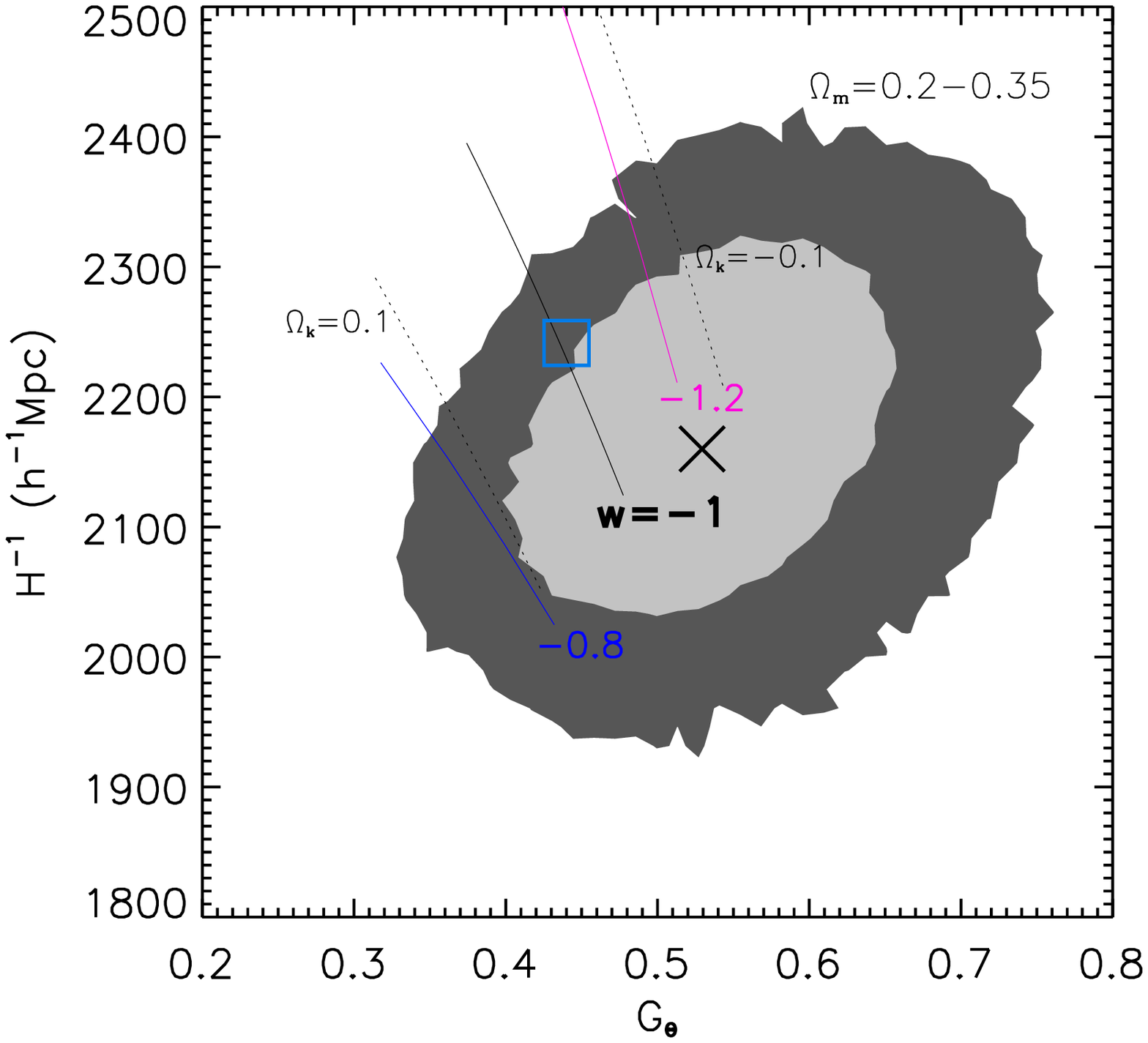}}
\resizebox{3.5in}{!}{\includegraphics {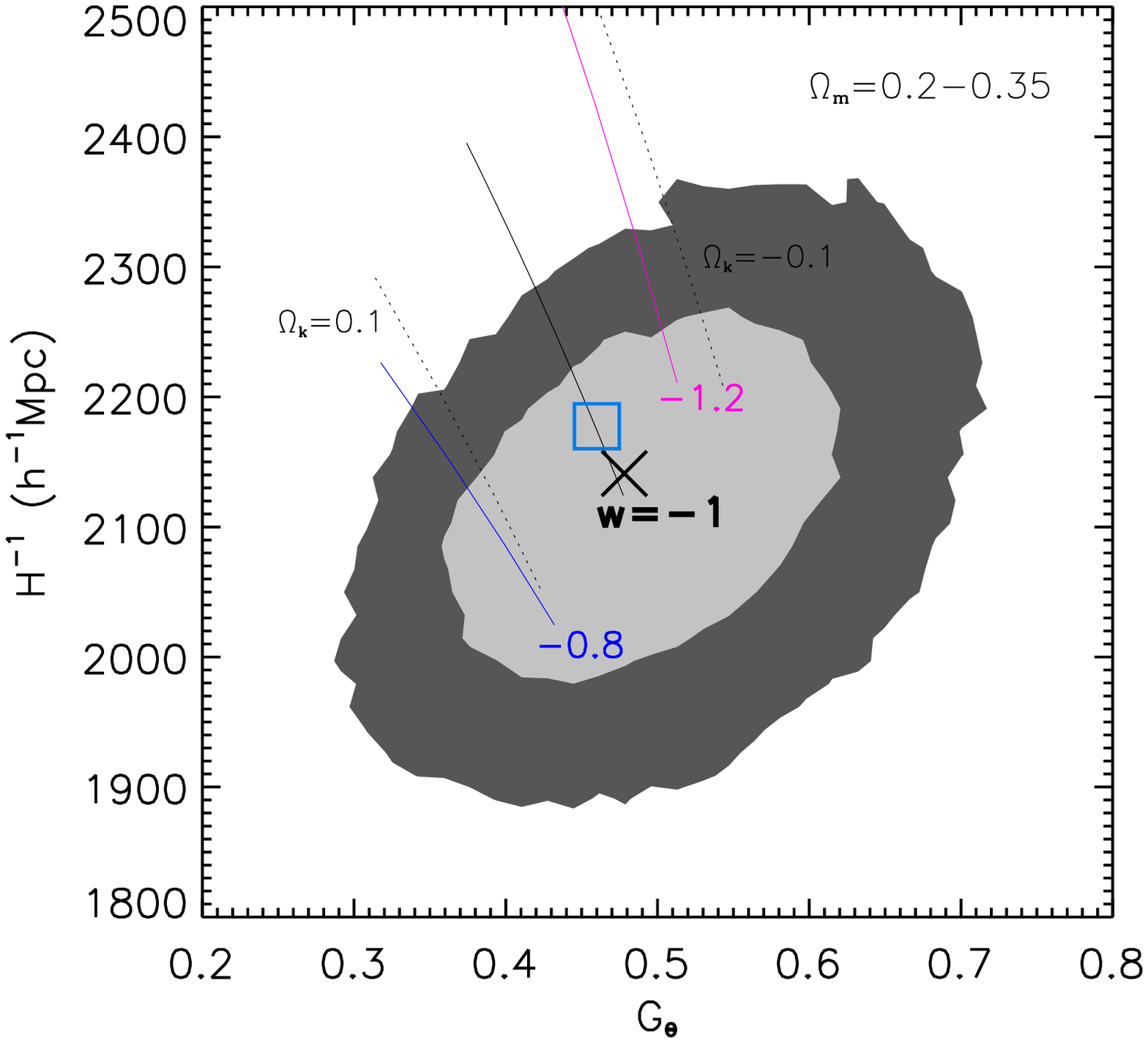}}
\end{center}
\caption{As Fig.~\ref{fig:dah} but for the $G_\theta-H^{-1}$ plane.  Here 
the allowed cosmology band is wider (we do not plot the owCDM models). 
} 
\label{fig:hg}
\end{figure*}

\section{Testing Cosmology} \label{sec:cosmology} 

Our analysis approach has been model independent, obtaining constraints 
on the distances $D_A$ and $H^{-1}$ -- without even assuming a Friedmann 
integral relation between them -- and on the velocity growth factor $G_\Theta$. 
While we have so far compared the values individually to the best fit LCDM 
predictions from the CMB, we should also look at the joint probabilities. 
We can test for consistency with the LCDM model by examining whether the 
fixed relations between these quantities in LCDM, i.e.\ the 1D curves in 
the $D_A-H^{-1}$, $D_A-G_\Theta$, and $H^{-1}-G_\Theta$ planes, 
all intersect the measured confidence contours.  Furthermore, we can 
generalize the test by allowing for spatial curvature or non-$\Lambda$ 
dark energy.  For the growth factor $G_\Theta$ this comparison also allows a 
test of general relativity since within this theory the distance quantities 
(measuring the cosmic expansion) have a definite relation to the growth 
quantity $\gth$. 

Figures~\ref{fig:dah}, \ref{fig:dag}, \ref{fig:hg} show the three planes 
of pairs of the cosmological quantities and their joint measurement contours, 
overlaid with the allowed theory curves of LCDM, oLCDM (with spatial 
curvature), wCDM (dark energy with constant equation of state ratio $w$), 
and owCDM.  Each one is shown for a WMAP9 (left panels) or Planck (right 
panels) early universe prior.  

In the $D_A-H^{-1}$ space, the cosmological models all lie within 
a narrow swath, somewhat separated from the best fit point in the 
Planck prior case. 
However, the 68\% confidence level contour of the measurements 
overlaps the LCDM model. 
In the $D_A-G_\Theta$ or $H^{-1}-G_\Theta$ planes, the standard cosmologies 
span a wider range of the space.  In both planes the measurements are 
consistent with LCDM at the 68\% confidenece level. There is no sign of 
significant deviation from LCDM in either distances or growth, and hence no 
sign of deviation from general relativity either. 

Note, however, that if we attempt to push the data by using data to smaller, 
non-linear  
scales, then we do find deviations.  In particular, $\gth$ rapidly becomes 
underestimated, with values of 0.42 for a cutoff at 
$\sigma_{\rm cut}=30\mpcoh$ and 0.34 for $\sigma_{\rm cut}=20\mpcoh$. 
However increasing $\sigma_{\rm cut}$ above $40\mpcoh$ does not change the 
result, indicating the value has converged. 
Had we included the smaller scales, we would have found that no cosmology 
(LCDM, oLCDM, wCDM) would have given good fits to the measurement contours. 
Moreover, we would have apparent evidence for a violation of general 
relativity. The apparent strong growth suppression in the measured growth rate 
$\gth=d\delta/d\ln a$ would yield an apparent gravitational growth index 
$\gamma$ \cite{2005PhRvD..72d3529L} of $\gamma\gtrsim0.7$, in contrast to 
the value 0.55 for general relativity. 

\begin{figure}
\begin{center}
\resizebox{3.4in}{!}{\includegraphics {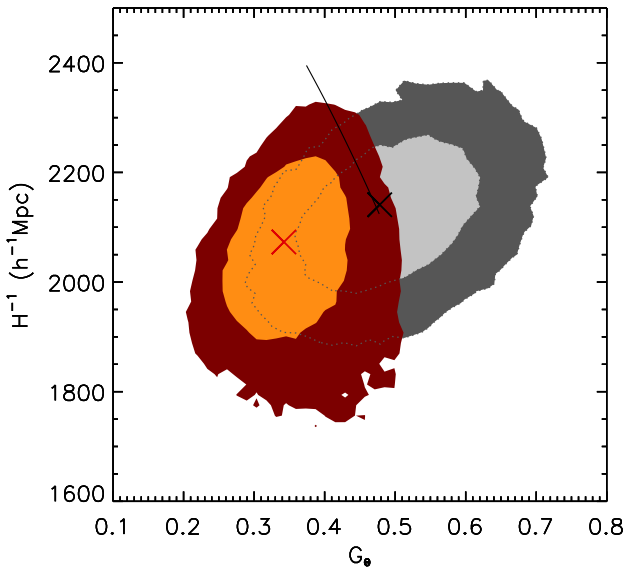}} 
\end{center}
\caption{Using small scale information, $\sigma_{\rm cut}=20\mpcoh$  
(in dark orange) rather than our standard $\sigma_{\rm cut}=40\mpcoh$ (in 
light grey), shifts the results into a 
region corresponding to no reasonable cosmology within general relativity. 
Here we show the $\gth-H^{-1}$ plane, with the Planck early universe prior, 
as an example. 
}
\label{fig:cut20}
\end{figure}

Figure~\ref{fig:cut20} shows what occurs in the cosmology parameters if 
data down to $\sigma_{\rm cut}=20\mpcoh$ is used.  The shifting of 
the best fit values, and the reduction in the uncertainty on $\gth$, 
clearly indicate that substantial information to fit 
cosmology is coming from small scales, not just the BAO ring scales. 
Unfortunately, the sensitivity of the results to low $\sigma_{\rm cut}$ 
(as opposed to the convergence found when 
$\sigma_{\rm cut}\gtrsim 40\mpcoh$) indicates that the modeling of the 
2D correlation function on these scales is inadequate.  Further improvements 
are necessary before these scales can be used to provide robust results. 

In terms of Fourier wavenumber, note that 
\be 
k\approx \frac{2\pi}{\sigma_{\rm cut}}= 
0.16\,\left(\frac{40\mpcoh}{\sigma_{\rm cut}}\right)\,h\,{\rm Mpc}^{-1} \ . 
\ee 
In comparisons to simulations the 2D anisotropic clustering model (not simply 
the 1D real space power spectrum or angle averaged correlation function) 
performed well down to $40\mpcoh$ \cite{Taruya:2010mx}.  
Another way to spuriously produce a shift outside the swath of standard 
cosmologies, and hence possibly conclude there is a violation of general 
relativity, is to misestimate $\zeff$. In fact, as we discuss in 
Appendix~\ref{sec:zeff}, $\zeff$ is itself anisotropic and will differ for 
different cosmological quantities but not at a level significant with current 
data.

\section{Conclusions} \label{sec:concl} 

We have used the BOSS CMASS DR9 galaxies to perform a cosmology model 
independent, fully 2D anistropic clustering analysis. Using an early 
universe prior from CMB experiments, from the clustering correlation 
function we can 
extract the angular diameter distance $D_A$, Hubble scale $H^{-1}$, and 
growth rate $\gth$ at the effective survey redshift $\zeff=0.57$. These 
are found to be consistent with LCDM, and by comparing expansion of 
cosmic distances with growth of cosmic structure we also test general 
relativity, again finding consistency. 

Two cautions are relevant to such an analysis, one important already to 
current data and one entering for future, high precision surveys. Use of 
small scale measurements of the correlation functions, which can be 
significantly contaminated by non--linear gravitational physics, is 
fraught with peril. We find this can distort the cosmological results, 
moving them wholly outside the range of standard cosmology and give 
a spurious signature of breakdown of general relativity. Insidiously, the 
extra data also helps shrink the contours, so the cosmological quantities 
appear well determined. 

We employ the improved redshift distortion model of \cite{Taruya:2010mx}, 
but this is still limited in accuracy to scales where higher order terms of 
the FoG effect are negligible. To prevent bias we cut most of the measured 
$\xi(\sigma,\pi)$ along the line of sight out from this analysis. This 
conservative treatment is well defined in the full 2D anisotropy analysis 
but could be problematic when using a multipole expansion instead. It will 
be interesting to compare our conservative results to those from a multipole 
analysis. 

Another aspect is that we find that the results from the real, observed, 
data are more contaminated with the small scale velocity and non-linear 
effects than those from the mock catalogues. In the simulations, 
$\sigma_{\rm cut}=20\mpcoh$ is acceptable to measure observables using the 
improved perturbation theory model. However, in the real dataset, the 
cut--off scale must be extended to $\sigma_{\rm cut}=40\hompc$ to obtain 
convergent results (insensitive to the exact choice of $\sigma_{\rm cut}$). 
This can also be seen by comparing the 2D BAO ring with the measured BAO peak 
structure. 

The second caution comes from the interpretation dependence of the 
effective redshift $\zeff$. Since it involves the galaxy power spectrum 
(or correlation function) it is intrinsically anisotropic and will take 
on different values depending on what quantity is being measured. That is, 
one formally has $D_A(\zeff^D)$, $H^{-1}(\zeff^H)$, etc. We estimate the 
magnitude of this effect and show that it could become relevant for next 
generation redshift surveys such as DESI or Euclid.

\acknowledgments 

Several authors thank KASI for hospitality during research visits. We 
especially thank Atsushi Taruya for many inputs and comments, and we 
thank Seokcheon Lee for reading the manuscript. 
This work was supported in part by the US DOE grant DE-SC-0007867 and 
Contract No.\ DE-AC02-05CH11231, and Korea WCU grant R32-10130 and Ewha 
Womans University research fund 1-2008-2935-001-2.  
Numerical calculations were performed by using a high
performance computing cluster in the Korea Astronomy and Space Science
Institute and we also thank the Korea Institute for Advanced Study for 
providing computing resources (KIAS Center for Advanced Computation Linux 
Cluster System).

\appendix 

\section{Effective redshift variation} \label{sec:zeff} 

The transverse and radial distances extracted from the galaxy data do not 
in fact have the same $\zeff$, as the optimal weighting depends on the strength 
of clustering \cite{1994ApJ...426...23F}, enhanced along the line of sight by redshift 
space distortions [e.g.\ the usual Kaiser factor $(b+f\mu^2)^2$].  
This is most familiar perhaps in the power spectrum, 
where the weighting $1/[1+n(z)\,P(k,\mu,z)]$ shows that the higher power 
along the line of sight further deweights lower redshift galaxies where 
clustering has grown. 

\begin{figure}
\begin{center} 
\resizebox{3.4in}{!}{\includegraphics{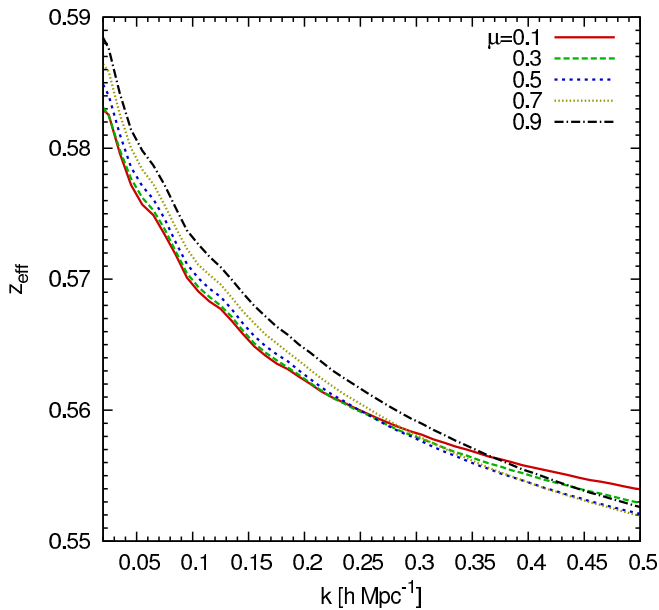}}
\end{center}
\caption{The effective redshift $\zeff$ of the galaxy sample depends 
on the clustering power and hence varies with wavenumber $k$ and redshift 
distortion angle $\mu$.  Observations with most probative power near the 
$k\approx 0.1\hompc$ have $\zeff\approx0.57$ for 
this data, but radial modes and transverse modes differ by 
$\Delta\zeff\approx 0.004$, potentially important for future surveys.} 
\label{fig:zeff}
\end{figure}

This is a small effect, negligible for previous redshift surveys, but will 
become increasingly important for larger, more precise surveys.  
Figure~\ref{fig:zeff} calculates $\zeff$ as a function of $k$ and $\mu$, 
using the power spectrum computed from mock simulations relevant to 
BOSS \cite{2012JCAP...11..014O}. 
Since most of the information for determining $H^{-1}$ comes from radial 
modes $\mu\approx 1$ and for determining $D_A$ comes from transverse modes 
$\mu\approx 0$, we see that the fit quantities are really $D_A(\zeff^D)$ and 
$H^{-1}(\zeff^H)$ where $\zeff^H-\zeff^D\approx 0.004$. 
This in turn would affect cosmological parameter estimation.

Around $\zeff\approx 0.57$, the Hubble parameter scales in 
LCDM as $H(z)\propto 1+z$ so that $dz/(1+z)\approx -dH^{-1}/H^{-1}$.  
Note that a survey that should use $\zeff=0.58$ rather than 0.57, say, 
due to the anisotropy 
of $\zeff$, would bias $H^{-1}$ by $-0.7\%$. This could be relevant for 
next generation surveys. Since $D_A$ is extracted mostly from the transverse 
modes where the observed clustering is equal to the real space clustering, 
no shift should be needed in the conventional $\zeff$ estimation.  (For 
completeness we note that if $\zeff=0.58$ rather than 0.57 then $D_A$ is 
biased high by 0.8\%.)  For $\gth$ the 2D anisotropy dependence is more 
complicated (see Fig.~4b of \cite{Song:2013ejh}).  
However, $G_\theta$ is near its maximum at $z=0.5$, so a change from 
$\zeff=0.57$ has a very small effect on it; in fact for $\zeff=0.58$ 
the bias is only $-0.0003$ or $-0.06\%$. 
Thus for current data precision the 
effect of different $\zeff$ for different cosmological parameters is 
negligible.  Next generation galaxy redshift surveys such as DESI or 
Euclid, however, should adapt $\zeff$ to the specific parameter being 
constrained.


\end{document}